\documentclass{article}

\usepackage[final,nonatbib]{nips_2017}

\usepackage[utf8]{inputenc} 
\usepackage[T1]{fontenc}    
\usepackage{hyperref}       
\usepackage{url}            
\usepackage{booktabs}       
\usepackage{amsfonts}       
\usepackage{nicefrac}       
\usepackage{microtype}      
\usepackage{graphicx}
\usepackage{caption}
\usepackage{subcaption}

\title{Time Domain Neural Audio Style Transfer}

\author{
  Parag K. Mital\\
  Kadenze, Inc.\thanks{http://kadenze.com}\\
  \texttt{parag@kadenze.com} \\
}

\begin{document}

\maketitle

\begin{abstract}
  A recently published method for audio style transfer has shown how to extend the process of image style transfer to audio.  This method synthesizes audio "content" and "style" independently using the magnitudes of a short time Fourier transform, shallow convolutional networks with randomly initialized filters, and iterative phase reconstruction with Griffin-Lim.  In this work, we explore whether it is possible to directly optimize a time domain audio signal, removing the process of phase reconstruction and opening up possibilities for real-time applications and higher quality syntheses.  We explore a variety of style transfer processes on neural networks that operate directly on time domain audio signals and demonstrate one such network capable of audio stylization.
\end{abstract}

\section{Introduction}


Audio style transfer \cite{Ulyanov2016} attempts to extend the technique of image style transfer \cite{Gatys} to the domain of audio, allowing "content" and "style" to be independently manipulated.  Ulyanov et al. demonstrates the process using the magnitudes of a short time Fourier transform representation of an audio signal as the input to a shallow untrained neural network, following similar work in image style transfer \cite{Ulyanov2016b}, storing the activations of the content and Gram activations of the style.  A noisy input short time magnitude spectra is then optimized such that its activations through the same network resemble the target content and style magnitude's activations.  The optimized magnitudes are then inverted back to an audio signal using an iterative Griffin-lim phase reconstruction process \cite{Griffin1984}.

Using phase reconstruction ultimately means the stylization process is not modeling the audio signal's fine temporal characteristics contained in its phase information.  For instance, if a particular content or style audio source were to contain information about vibrato or the spatial movement or position of the audio source, this would likely be lost in a magnitude-only representation.  Further, by relying on phase reconstruction, some error during the phase reconstruction is likely to happen, and developing real-time applications are also more difficult \cite{Wyse2017}, though not impossible \cite{Prusa2017}.  In any case, any networks which discard phase information, such as \cite{Wyse2017}, which build on Ulyanov's approach, or recent audio networks such as \cite{Hershey2016} will still require phase reconstruction for stylization/synthesis applications.

Rather than approach stylization/synthesis via phase reconstruction, this work attempts to directly optimize a raw audio signal.  Recent work in Neural Audio Synthesis has shown it is possible to take as input a raw audio signal and apply blending of musical notes in a neural embedding space on a trained WaveNet autoencoder \cite{Engel2017}.  Though their work is capable of synthesizing raw audio from its embedding space, there is no separation of content and style using this approach, and thus they cannot be independently manipulated.  However, to date, it is not clear whether this network's encoder or decoder could be used for audio stylization using the approach of Ulyanov/Gatys.

To understand better whether it is possible to perform audio stylization in the time domain, we investigate a variety of networks which take a time domain audio signal as input to their network: using the real and imaginary components of a Discrete Fourier Transform (DFT); using the magnitude and unwrapped phase differential components of a DFT; using combinations of real, imaginary, magnitude, and phase components; using the activations of a pre-trained WaveNet decoder \cite{Oord2016b,Engel2017}; and using the activations of a pre-trained NSynth encoder \cite{Engel2017}.  We then apply audio stylization similarly to Ulyanov using a variety of parameters and report our results.

\section{Experiments}

We explore a variety of computational graphs which use as their first operation a discrete Fourier transform in order to project an audio signal into its real and imaginary components.  We then explore manipulations on these components, including directly applying convolutional layers, or undergoing an additional transformation of the typical magnitude and phase components, as well as combinations of each these components.  For representing phase, we also explored using the original phase, the phase differential, and the unwrapped phase differentials.  From here, we apply the same techniques for stylization as described in \cite{Ulyanov2016}, except we no longer have to optimize a noisy magnitude input, and can instead optimize a time domain signal.  We also explore combinations of using content/style layers following the initial projections and after fully connected layers.

We also explore two pre-trained networks: a pre-trained WaveNet decoder, and the encoder portion of an NSynth network as provided by Magenta \cite{Engel2017}, and look at the activations of each of these networks at different layers, much like the original image style networks did with VGG.  We also include Ulyanov's original network as a baseline, and report our results as seen through spectrograms and through listening.  Our code is also available online\footnote{https://github.com/pkmital/neural-audio-style-transfer}\footnote{Further details are described in the Supplementary Materials}.

\section{Results}

Only one network was capable of producing meaningful audio reconstruction through a stylization process where both the style and content appeared to be retained: including the real, imaginary, and magnitude information as concatenated features in height and using a kernel size 3 height convolutional filter.  This process also includes a content layer which includes the concatenated features before any linear layer, and a style layer which is simply the magnitudes, and then uses a content and style layer following each nonlinearity.  This network produces distinctly different stylizations to Ulyanov's original network, despite having similar parameters, often including quicker and busier temporal changes in content and style.  The stylization also tends to produce what seems like higher fidelity syntheses, especially in lower frequencies, despite having the same sample rate.  Lastly, this approach also tends to produce much less noise than Ulyanov's approach, most likely due to errors in the phase reconstruction/lack of phase representation.

Every other combination of input manipulations we tried tended towards a white noise signal and did not appear to drop in loss.  The only other network that appeared to produce something recognizable, though with considerable noise was using the magnitude and unwrapped phase differential information with a kernel size 2 height convolutional filter.  We could not manage to stylize any meaningful sounding synthesis using the activations in a WaveNet decoder or NSynth encoder.


\section{Discussion and Conclusion}

This work explores neural audio style transfer of a time domain audio signal.  Of these networks, only two produced any meaningful results: the magnitude and unwrapped phase network, which produced distinctly noisier syntheses, and the real, imaginary, and magnitude network which was capable of resembling both the content and style sources in a similar quality to Ulyanov's original approach, though with interesting differences.  It was especially surprising that we were unable to stylize with NSynth's encoder or decoder, though this is perhaps to due to the limited number of combinations of layers and possible activations we explored, and is worth exploring more in the future.



\small

\begin{figure}
\centering
\includegraphics[width=1\linewidth]{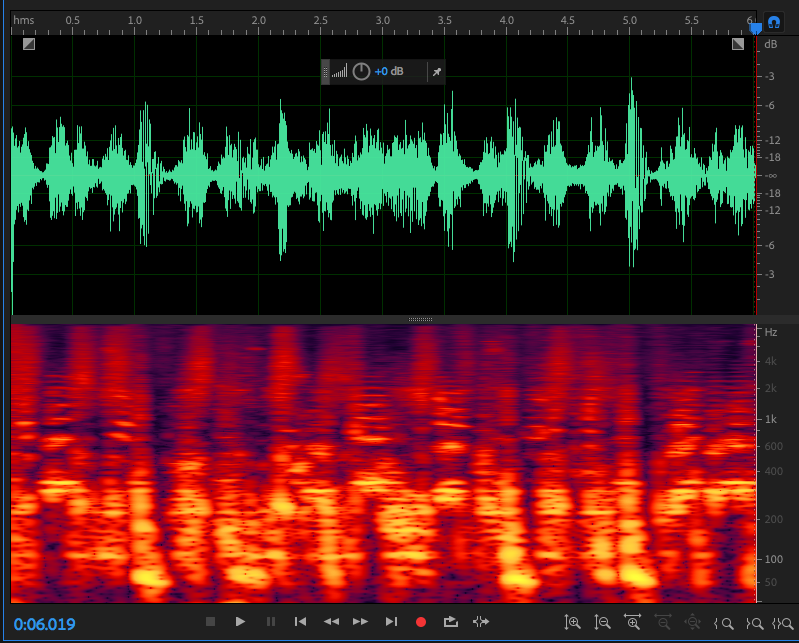}
\caption{Example synthesis optimizing audio directly with both the source content and style audible.}
\end{figure}
\end{document}